\documentclass[oribibl]{llncs}

\PassOptionsToPackage{hyphens}{url}
\usepackage[colorlinks = true,
  linkcolor = blue,
  citecolor = blue,
  urlcolor = blue]{hyperref}

\usepackage{url}
\usepackage{csquotes}
\usepackage{multirow}
\usepackage{amsmath}
\usepackage{amssymb}
\usepackage{graphicx}
\usepackage{tabularx}
\usepackage{booktabs}
\usepackage{threeparttable}

\begin{document}

\title{A Rapid Review Regarding the Concept of Legal Requirements in
  Requirements Engineering}

\author{Jukka Ruohonen\orcidID{0000-0001-5147-3084} \\ \email{juk@mmmi.sdu.dk}}
\institute{University of Southern Denmark, S\o{}nderborg, Denmark}

\maketitle

\begin{abstract}
Out of a personal puzzlement, recent peer review comments, and demonstrable
confusion in the existing literature, the paper presents a rapid review of the
concept of legal requirements (LRs) in requirements engineering (RE)
research. According to reviewing results, a normative understanding of LRs has
often been present, although proper definitions and conceptual
operationalizations are lacking. Some papers also see LRs as functional and
others as non-functional requirements. Legal requirements are often
characterized as being vague and complex, requiring a lot of effort to elicit,
implement, and validate. These characterizations supposedly correlate with
knowledge gaps among requirements engineers. LRs are also seen to often change
and overlap. They may be also prioritized. According to the literature, they
seem to be also reluctantly implemented, often providing only a minimal baseline
for other requirements. With these and other observations, the review raises
critical arguments about apparent knowledge gaps, including a lack of empirical
evidence backing the observations and enduring conceptual confusion.
\end{abstract}


\section{Introduction}

Legal requirements are a classical topic in RE research. Yet, there is arguably
a widespread confusion about them in the existing literature---or at least the
author of this rapid review has encountered the confusion and being confused by
himself. How to characterize LRs? When contemplating an answer, together with
colleagues we recently considered the separation between functional and
non-functional requirements, and ended to characterize LRs as
non-negotiable~\cite{Ruohonen25ESPREb}. On a second though, this
characterization was probably a bad choice because we have also emphasized
argumentative RE with respect to LRs and a need to potentially demonstrate
compliance~\cite{Hjerppe19RE, Ruohonen25ISTb}. This personal reflection
motivates the paper's rapid review about whether similar confusion has been met
by others.

Without further ado, in what follows, three research questions are examined. The
first research question (RQ.1) is: \textit{what legal requirements are according
  to the existing literature?} More than anything, the question is
philosophical; it is about the perceptions and opinions RE researchers have had
about law through LRs. By mimicking ontology-building research for
LRs~\cite{Abualhaija24}, the subsequent RQ.2 asks: \textit{what characteristics
  legal requirements have according to the existing literature?} If RQ.1 is
about the nature of LRs, RQ.2 seeks to ask about the qualifying characteristics
of LRs. Such characteristics reflect the different qualifying nouns, verbs, and
adjectives often present in laws~\cite{Abualhaija24, Ruohonen25ISTb}. The last
RQ.3 is: \textit{how legal requirements affect the work of requirements
  engineers beyond elicitation?} As will become clear in the subsequent section
on the materials and methods, the answers given to these questions are only
tentative. Nevertheless---even with tentative answers, research gaps will be
demonstrated and critical arguments~raised.

\section{Materials and Methods}

The paper is a \textit{rapid review} of literature. A rapid review has been
defined as ``\textit{a~form of knowledge synthesis that accelerates the process
  of conducting a traditional systematic review through streamlining or omitting
  a variety of methods to produce evidence in a resource-efficient manner}''
\cite[p.~80]{Hamela21}. Thus, efficiency and feasibility are the driving forces
behind rapid reviews. In addition, a rapid review is suitable for the present
work because none of the three RQs deal with empirical or other
\textit{evidence} (cf.~\cite{Song21}), which has been seen as the qualifying
noun for systematic literature reviews~\cite{Kitchenham09}. Therefore, neither
is there any cumulation of evidence; the answers to the RQs do not become
stronger the more there are papers raising the same points. That said, the
efficiency and feasibility goalposts necessitate a limitation that not all
arguments and points are covered. Despite this limitation, a systematic approach
was still used for sampling literature.

The literature sample is based on Scopus. With feasibility in mind, the sampling
was started on 6 September 2025 by using the keywords ``\textit{legal
  requirement}'' and ``\textit{requirements engineering}'', separated by the
Boolean AND operator and including the quotation marks. When a query was
conducted from all fields of papers written in English, $602$ papers were
returned. This amount is infeasible already because downloading so many papers
from publishers' closed online databases would take a lot of time. Therefore, to
reduce the amount, only open access papers were first qualified. With this
restriction, $154$ papers were returned. These were selected as the initial
sample, which was further reduced by excluding papers that offered neither a
definition for legal requirements nor a characterization of these. To clarify:
those papers were excluded that operated with specific laws and their
requirements without providing a conceptual, theoretical, or legal definition or
characterization for the concept of legal requirements in general. In practice,
the exclusions and inclusions were done by searching the keyword ``\textit{legal
  requirement}'' and reading the sentences in which the keyword
appeared.\footnote{~In addition, as many as $68$ papers had to be excluded
because these were not actually provided as (gold or diamond) open access even
though Scopus claimed so. This point serves to remind that scientific meta-data
is riddled with problems.}

With these restrictions, only $21$ papers were qualified into the
sample. Although feasibility was satisfied, the sample size is certainly on the
small side. Therefore, two additional points are warranted. The first point is
that the sample contains two existing and exhaustive literature reviews \cite{2,
  9}, but neither one addressed the RQs specified. Therefore, it can be argued
that even the twenty-one papers sampled are sufficient for saying something
new. The second point is that the systematically sampled literature is augmented
with additional, subjectively selected literature about LRs. Analogously to
existing reviews~\cite{Song21}, the rapid review thus mixes systematic sampling
with a ``classical'' literature review approach, perhaps thus being a
``semi-systematic rapid review'' in overall. Regarding presentation of the
results, a thematic approach is used partially again due to its feasibility for
summarizing papers with concisely labeled categories.

\section{Results}

The themes identified for the three research questions are enumerated in
Table~\ref{tab: themes}. Before continuing to elaborate these, it should be
noted that the papers listed in the table are what was retrieved through the
systematic sampling; anything that is referenced but does not appear in the
table is about the ``classical''~part.

\begin{table}[th!b]
\centering
\caption{The Themes Identified}
\label{tab: themes}
\begin{tabular}{llcllcll}
\toprule
\multicolumn{2}{c}{RQ.1} && \multicolumn{2}{c}{RQ.2} && \multicolumn{2}{c}{RQ.3} \\
\hline
1. & Normative && 1. & Ambiguities && 1. & Effort \\
& \cite{6, 8, 9} &&& \cite{1, 18, 7, 4, 20, 17, 3} &&& \cite{5, 8} \\
\cmidrule{2-2}\cmidrule{5-5}\cmidrule{8-8}
2. & Functional \cite{5, 15} && 2. & Baselines \cite{18} && 2. & Expectations \cite{6, 7} \\
\cmidrule{2-2}\cmidrule{5-5}\cmidrule{8-8}
&&& 3.& Benefits \cite{7, 14} && 3. & Knowledge  \cite{10, 7, 8, 16} \\
\cmidrule{5-5}\cmidrule{8-8}
&&& 4. & Changes \cite{1, 18} && 4. & Prioritization \cite{7} \\
\cmidrule{5-5}\cmidrule{8-8}
&&& 5. & Complexities && 5. & ``Reluctance'' \\
&&&    & \cite{5, 7, 8, 4, 2, 12} &&& \cite{13, 7, 8} \\
\cmidrule{5-5}\cmidrule{8-8}
&&& 6. & Conflicts \cite{3} && 6. & Timing \cite{13, 21} \\
\cmidrule{5-5}\cmidrule{8-8}
&&& 7. & Documents \cite{12} && 7. & ``Validation'' \cite{19, 8, 4, 11, 9}  \\
\cmidrule{5-5}\cmidrule{8-8}
&&& 8. & Overlaps \cite{7, 8, 17, 12} && \\
\bottomrule
\end{tabular}
\end{table}

The reviewing is best done through going through the three RQs
consecutively. Thus, regarding RQ.1, only a few studies contemplated about the
nature of LRs. When contemplations were present, a normative understanding was
often present too. For instance, a paper elaborated LRs by stating that
justifications for these are based on ``\textit{external norms}''
\cite[p.~2]{6}. Although no definition was provided for such external norms,
these can be seen to reflect a distinction between regulative norms dealing with
deontic modalities (such as explicit prohibitions or permissions) and
constitutive norms dealing with declarations and definitions for existing
facts~\cite{Abualhaija24}. Given this distinction, a paper stated that legal
``\textit{requirements are extracted and annotated using deontic logic for
  obligations and permissions}'' \cite[p.~345]{9}. The paper omitted
constitutive norms, while another paper omitted deontic modalities by stating
that it dealt with ``\textit{the translation of real-world facts, which we
  define as the construct legal requirements}'' \cite[p.~2322]{8}. These two
quotations support an argument that the distinction should be better taken into
account in future research. A related point is about the enduring confusion and
debates regarding framing of LRs as either functional or non-functional
requirements~\cite{Ruohonen25ESPREb}. Regarding this confusing debate, a couple
of papers leaned toward the functional requirement camp. For instance, LRs were
seen to \textit{describe the behavior and functions of a software system}''
\cite[p.~2]{5}. Analogously, ``\textit{research on legal requirements mainly
  focuses on the behavior of a machine learning model}'' \cite[p.~20]{15}. The
nouns behavior and functions align well with the deontic modalities that may
allow or prohibit a certain behavior of a system.

Regarding RQ.2, LRs were frequently seen to contain different ambiguities and
vagueness in general. Many ``\textit{broader ambiguities}'' were argued to
characterize LRs with respect to how they are~``\textit{interpreted and
  applied}'' \cite[p.~11]{3}. Analogously, ``\textit{legal requirements are
  often vaguely formulated and leave room for interpretation}''
\cite[p.~446]{7}. This ambiguity term correlated with a frequently occurring
complexity theme. As summarized in an existing literature review, the RE
research literature often perceives the complexity them to have a causal logic:
LRs are complex because laws are complex~\cite[p.~12]{2}. By hypothesis, the
ambiguity and complexity themes are related to, or correlated with, a third
theme identified as a partial answer to RQ.3. This theme is about the knowledge
and skills requirements engineers, or engineers in general, have---or, rather,
do not have. The following quotation summarizes the frequent knowledge theme
well: legal requirements ``\textit{increase pressure on developers of systems
  who often lack the necessary domain expertise to implement regulation rules}''
\cite[p.~442]{7}. The onus was seen to be on the engineering side because it was
also stated that ``\textit{legal requirements are well documented}''
\cite[p.~1]{12}. Therefore, together the ambiguity, complexity, and knowledge
themes were frequently used to motivate a paper, often either in a form of a new
tool or technique proposed and implemented or an interdisciplinary approach
involving scholars of law. However, none of the arguments raised in the
literature sampled were based on empirical evidence.

To continue with RQ.2, the literature recognized that LRs often change due to
changes in laws and guidelines for complying with them~\cite{Hjerppe19RE,
  Ruohonen25ESPREb}. In short: ``\textit{legal requirements may change}''
\cite[p.~3]{1}. Also different overlaps were frequently recognized in the
systematically sampled literature, which aligns with the non-sampled
literature~\cite{Ruohonen25ESPREb, Ruohonen25ISTb}, including with respect to
overlapping and possibly conflicting laws in different jurisdictions~\cite{17,
  NegriRibalta25}. However, the change and overlap themes are not clear-cut
because it was also claimed that ``\textit{legal requirements stabilise over
  time and occur repeatedly in many development contexts in a similar form}''
\cite[p.~442]{7}. With this quotation in mind, it could be hypothesized that
perhaps a slow but still possible stabilization trend characterizes LRs despite
of changes and overlaps in laws and LRs derived from them. This hypothesis
aligns with an argument expressed in the literature that---despite of the
ambiguity, complexity, knowledge, change, and overlap themes---legal
requirements still often only provide a baseline. In other words,
``\textit{requirements of a system may go beyond legal requirements}''
\cite[p.~3]{18}, including with respect to non-legal security, privacy, and data
protection requirements~\cite{Hjerppe19RE}. Indeed, some new legal cyber
security requirements in Europe have been perceived as
sensible~\cite{Ruohonen25ESPREb}, although they are likely only minimal cyber
security requirements for some customers.

It is also important to emphasize that the negative connotations associated with
the ambiguity, complexity, knowledge, and related themes were balanced by
arguments that LRs have also~benefits. In other words, ``\textit{legal
  requirements are considered to be necessary for market approval, but the
  benefits and added value are often not understood}''
\cite[p.~448]{7}. Although a more thorough literature review would be required
to know whether there really is a knowledge gap, the quotation can be still used
to tentatively argue that more business research is needed for better
understanding potential benefits from LRs and compliance with laws in
general. Certification, including cyber security certification
\cite{Ruohonen25ISTb}, would be a good example in this regard. Finally,
regarding RQ.2, a paper dealing with blockchains argued that ``\textit{technical
  structure and governance models often clash with established legal
  requirements}'' \cite[p.~11]{3}. The quotation is a good remainder about the
general difficulties in marrying law and technology.

Regarding RQ.3, the knowledge theme for it correlates with an effort them; LRs
are ``\textit{a daunting task for requirements engineers}''
\cite[p.~2]{5}. Again, the claim quoted was provided without evidence. Regarding
the earlier baseline theme, there is also a ``reluctance theme'' (for a lack of
a better term). Despite other non-legal requirements, LRs are allegedly
``\textit{often only addressed to a minimum extent in order to be compliant}''
\cite[p.~2313]{8}. Furthermore, they are ``\textit{often a necessary burden for
  developers but one that is implemented rather reluctantly and with little
  effort in the end}'' \cite[p.~449]{7}. A possible explanation may relate to a
timing theme in a sense that LRs are supposedly ``\textit{often considered only
  once a protype has already been built}'' \cite[p.~388]{13}. If true, such an
\textit{ex~post} approach conflicts with many guidelines and design principles,
among them the notion of ``building security in''~\cite{McGraw04}. Furthermore,
there is a closely related theme about expectations; requirements engineers and
other engineers allegedly design and implement ``\textit{legal requirements
  supervisors expect to see}'' \cite[p.~5]{6}. This expectation theme is related
to a further prioritization theme according to which legal requirements
``\textit{may be prioritised higher due to their relevance and the risk of
  penalties}'' \cite[p.~445]{7}. This quote exemplifies a criticism that cyber
security and related risk analyses are supposedly often done with compliance in
mind, not in terms of what is being protected~\cite{Ruohonen25ISTb}. The last
theme is about~``validation''.

The ``validation'' theme has quotation marks because it remains unclear and
debatable how LRs should be validated. The reason is that these are
\textit{legal} requirements and thus validation, including an interpretation
involving ambiguities, will be done by regulators and possibly then ultimately
by courts. With this point in mind, there are papers dealing with
traceability~\cite{9} and evaluation of already compliant systems or their
components~\cite{19, 21}, among other things. In other words,
``\textit{specifying, monitoring, and testing software systems for compliance
  with legal requirements}'' \cite[p.~2]{4}, even in case only minimal baselines
are reluctantly implemented with supervisory expectations in mind. Whether
actual systematic monitoring and testing occurs in practice is another
question. Implicitly reflecting the term checkbox
exercises~\cite{Ruohonen25ISTb}, a paper argued that ``\textit{standards to
  support legal experts in tracking legal requirements by assessing which
  requirements of the law have been met}'' \cite[p.~2314]{8} are important. Even
with standards, ``\textit{the difficulty of compliance can be influenced by the
  complexity of legal requirements and the ability to measure compliance}''
\cite[p.~2]{4}. This quote serves well to end the ``semi-systematic'' reviewing
by emphasizing that not only do laws and legal requirements overlap but so do
also the themes in Table~\ref{tab: themes}.

\section{Conclusion}

The paper presented a rapid review on LRs in RE research. According to the
results, there indeed is a conceptual and theoretical confusion regarding the
nature of legal requirements. They are normative because they are about law but
they are still not about law only; they are also about functionalities and
behaviors of systems. The point about ``validation'' exemplifies this dual
law-technology nature of LRs. Even with the limited amount of literature
reviewed, a critical argument can be raised about a need to have a better and
preferably uniform conceptual and theoretical framework for LRs. Even having a
common terminology would be beneficial for RE research. The second critical
argument is about a seeming lack of robust empirical evidence regarding what
allegedly characterizes LRs and what requirements engineering are doing with
them.

As it stands, the situation seems to reflect the old argument about software
engineering sometimes being ``folklore turned into
facts''~\cite{Fernandez19}. Are LRs implemented only minimally and reluctantly?
Do engineers lack legal knowledge? If so, should improvements be sought through
better education rather than implementation of more and more tools and
frameworks? As these questions seem to lack empirical answers, and as none of
what is present in Table~\ref{tab: themes} is backed by robust empirical
evidence, surveying industry practitioners would be a good start. But it is not
only about the industry. Are laws really complex and ambiguous? While surveys
could be used to answer also to this question, it is also possible to measure
laws just like it is possible to measure software. In addition to classical
metrics such as entropy, readily available metrics include those related to
readability~\cite{Ruohonen21ICEDEG} and references within and across
laws~\cite{Sannier17}. Thus, the ongoing ontology-building
research~\cite{Abualhaija24} can be continued from a slightly different~angle.

\bibliographystyle{splncs03}

\end{document}